\documentclass[aps,prd,twocolumn,superscriptaddress,showpacs,preprintnumbers]{revtex4}

\usepackage{graphicx}
\usepackage{hyperref}

\newcommand{\s}{\hbox{ s}}	 
\newcommand{\y}{\hbox{ y}}	 
\newcommand{\ev}{\hbox{ eV}}

\newcommand{\mev}{\hbox{ MeV}}
\newcommand{\gev}{\hbox{ GeV}}

\newcommand{\nb}{\hbox{ nb}}
\newcommand{\cm}{\hbox{ cm}}
\newcommand{\km}{\hbox{ km}}
\newcommand{\mpc}{\hbox{ Mpc}}
\newcommand{\kpc}{\hbox{ kpc}}

\newcommand{\g}{\hbox{ g}}
\newcommand{\snu}{\ensuremath{\tilde{\nu}}}
\newcommand{\neut}{\ensuremath{\tilde{\chi}^{0}_{1}}}
\newcommand{\neuter}{\ensuremath{\tilde{\chi}}}

\begin{document}

\preprint{\textsf{FERMILAB--Pub--06/050--T}}

\title{Neutrino Coannihilation on Dark-Matter Relics?}


\author{Gabriela Barenboim}
\email[E-mail: ]{Gabriela.Barenboim@uv.es}
\affiliation{Departament de F\'{\i}sica Te\`{o}rica,
 Universitat de Val\`{e}ncia, 
Carrer Dr.~Moliner 50, E-46100 Burjassot (Val\`{e}ncia),  ÊÊSpain}
\author{Olga Mena Requejo}
\email[E-mail: ]{omena@fnal.gov}
\affiliation{Theoretical Physics Department\\ Fermi National 
Accelerator Laboratory\\ P.O.\ Box 500, Batavia, Illinois 60510}
\author{Chris Quigg}
\email[E-mail: ]{quigg@fnal.gov}
\affiliation{Theoretical Physics Department\\ Fermi National
Accelerator Laboratory\\ P.O.\ Box 500, Batavia, Illinois 60510}

\date{\today}

\begin{abstract}
High-energy neutrinos may resonate with relic background neutralinos to
form short-lived sneutrinos.  In some circumstances, the decay chain
that leads back to the lightest supersymmetric particle would yield
few-GeV gamma rays or charged-particle signals.  Although resonant
coannihilation would occur at an appreciable rate in our galaxy, the
signal in any foreseeable detector is unobservably small.
\end{abstract}

\pacs{96.50.Zc,95.35.+d,14.80.Ly}

\maketitle

The possibility of detecting relic neutrinos by observing the 
resonant annihilation of extremely energetic neutrinos on the 
background neutrinos through the reaction $\nu \bar{\nu} \to Z^{0}$ 
has been the object of extensive 
studies~\cite{Weiler:1982qy,Eberle:2004ua,Barenboim:2004di}.  Given 
the small neutrino masses ($\lesssim 1\ev$) indicated by current 
experimental constraints, suitably intense sources of extremely energetic 
($10^{21}$ -- $10^{25}$-eV) cosmic neutrinos are required. The 
positions and shapes of the absorption lines in the extremely high-energy 
neutrinos spectrum are influenced by Fermi motion of the relics and by 
the thermal history of the Universe.

Relic neutrinos have a special standing: According to the standard
cosmology, neutrinos should be the most abundant particles in the
Universe, after the photons of the cosmic microwave background,
provided that they are stable over cosmological times.  But the weight
of cosmological observations argues that neutrinos are not the only
undetected relics.  According to the Wilkinson Microwave Anisotropy
Probe (WMAP) three-year analysis~\cite{Spergel:2006hy}, the matter
fraction of the present Universe is $\Omega_{m}h^{2} =
0.127^{+0.007}_{-0.014}$, where $h = 0.73 \pm 0.03$ is the reduced
Hubble parameter~\cite{PDGcosmo} and $\Omega_{m}$ is the ratio of the
matter density to the critical density $\varrho_{c} \equiv
{3H^{2}}/{8\pi G_{\mathrm{N}}}$.  (Here $H$ is the Hubble parameter and
$G_{\mathrm{N}}$ is Newton's constant.)  The baryonic fraction is
$\Omega_{b}h^{2} = 0.0223^{+0.0007}_{-0.0009}$, and the neutrino
fraction $\Omega_{\nu}h^{2} = \left(\sum_{i} m_{\nu_{i}}\right)/94\ev
\approx 0.0072 \lesssim \Omega_{b}h^{2}$.  Accordingly, we have reason
to believe that the matter fraction is dominated by an unseen
``dark-matter'' component.  A popular hypothesis, realized in many 
extensions to the standard model including
supersymmetry, holds that the dark matter is dominated by a
single species of weakly interacting massive particle (WIMP).

There is no confirmed observation of the passage of WIMPs through a
detector~\cite{Bertone:2004pz,Eidelman:2004wy}.  A positive signal
reported by the dark matter experiment DAMA~\cite{Bernabei:2003za}
seems in conflict with upper limits set by the Cryogenic Dark Matter
Search~\cite{Akerib:2005kh} and the EDELWEISS
experiment~\cite{Sanglard:2005we}.  Accelerator experiments have so far
not yielded evidence for the production of a superpartner candidate for
the dark-matter particle~\cite{Eidelman:2004wy}.

\begin{figure}[tb]
   \centerline{\includegraphics[width=3.0cm]{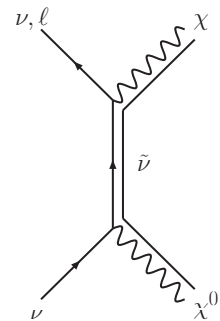}}
   \vspace*{-12pt}
\caption{Resonant sneutrino formation in neutrino--neutralino 
collisions. Double lines denote superpartners.}
\label{fig:feynman}
\end{figure}

In this note, we ask whether resonant coannihilation of neutrinos with
dark-matter superpartners might be observable.  We study a particular
representative case of neutralino dark matter, in which the absolutely
stable lightest supersymmetric particle (LSP) is $\neut$, a
superposition of neutral wino, bino, and Higgsino.  In circumstances
apt for detection, resonant formation of a sneutrino in the reaction
$\nu \neut \to \tilde{\nu}$, shown in Figure~\ref{fig:feynman},
might
induce absorption lines or direct 
signatures.  A  sneutrino heavier than the
neutralino is implied by the assumption that $\neut$ be the
LSP. A direct signature of sneutrino formation and decay requires 
that \snu\  decay into channels beyond the $\nu\neut$ formation
channel.  Informative examples include the parameter sets (I' and
L') presented among the post-WMAP
benchmarks for the constrained minimal supersymmetric standard 
model in Ref.~\cite{Battaglia:2003ab}.  (The same two models have been considered by Datta and 
collaborators~\cite{Datta:2004sr} in their recent study of the 
Lorentz-boosted situation of ultrahigh-energy neutralinos scattering 
on the relic neutrino background.)

We summarize the relevant information in 
Table~\ref{table:sugra}.
\begin{table}
\caption{Superpartner parameters in two MSSM
scenarios~\cite{Battaglia:2003ab}, evaluated using the
\textsf{SPheno}~\cite{Porod:2003um} code for particle properties and
the \textsf{MicrOMEGAs}~\cite{Belanger:2004yn} code for relic
densities, as implemented at the \textit{Comparison of SUSY spectrum generators}
web site~\cite{Belanger:2005jk} with $m_{t} = 
172.5\gev$~\cite{Group:2006qt} and the default values $m_{b} = 
4.214\gev$ and $\alpha_{s}(M_{Z}^{2}) =0.1172$.\label{table:sugra} }
\begin{ruledtabular}
    \begin{tabular}{c c c}
 & Model I' & Model L' \\
 \hline
 $M(\neut)$ & $141\gev$ & $185\gev$ \\
 $\Omega_{\tilde{\chi}}h^{2}$ & $0.105$ & $0.111$ \\
 {$\bar{\mathcal{N}}_{\tilde{\chi}}$} & {$7.7 \times 10^{-9}\cm^{-3}$} & 
 {$6.4 \times 10^{-9}\cm^{-3}$} \\
 {$\mathcal{N}_{\tilde{\chi}}(r_{\odot})$} & {$2.2 \times 
 10^{-3}\cm^{-3}$} & 
 {$1.6 \times 10^{-3}\cm^{-3}$} \\
 {$\mathcal{N}_{\tilde{\chi}}^{\oplus}$} & {$< 400\cm^{-3}$} & {$< 300\cm^{-3}$} \\[3pt]
 $M(\tilde{\chi}^{0}_{2})$ & $265\gev$ & $349\gev$ \\
 $M(\tilde{\chi}^{\pm})$ & $265\gev$ & $349\gev$ \\[3pt]
 $m_{\tilde{\nu}_{e,\mu}}$ & $290\gev$ & $422\gev$ \\
 {$E_{\nu}^{\mathrm{res}}$} & {$299\gev$} & {$481\gev$} \\
 {$\Gamma_{\tilde{\nu}}$} & {$324\mev$} & {$857\mev$} \\
{$\sigma^{\mathrm{peak}}_{\tilde{\nu}:\mathrm{inel}}$} & $40.2\nb$ & 
$20.6\nb$ \\
{$\Gamma_{\tilde{\nu}}\,\sigma^{\mathrm{peak}}_{\tilde{\nu}:\mathrm{inel}}$} & 
$13.0\nb\gev$ & $17.7\nb\gev$ \\[3pt]
$\mathcal{B}(\snu \to \neut \nu)$ & $0.720$ & $0.434$ \\
$\mathcal{B}(\snu \to \tilde{\chi}^{0}_{2} \nu)$ & $0.086$ & $0.179$ \\
$\mathcal{B}(\snu \to \tilde{\chi}^{\pm} \ell^{\mp})$ & $0.194$ & $0.386$ \\[6pt]
$m_{\tilde{\nu}_{\tau}}$ & $277\gev$ & $386\gev$ \\
{$E_{\nu}^{\mathrm{res}}$} & {$272\gev$} & {$403\gev$} \\
{$\Gamma_{\tilde{\nu}}$} & {$612\mev$} & {$2.011\gev$} \\
{$\sigma^{\mathrm{peak}}_{\tilde{\nu}:\mathrm{inel}}$} & {$52.3\nb$} & 
{$14.3\nb$} \\
{$\Gamma_{\tilde{\nu}}\,\sigma^{\mathrm{peak}}_{\tilde{\nu}:\mathrm{inel}}$} & 
{$32.0\nb\gev$} & {$28.8\nb\gev$} \\[3pt]
$\mathcal{B}(\snu \to \neut \nu_{\tau})$ & $0.342$ & $0.153$ \\
$\mathcal{B}(\snu \to \tilde{\chi}^{0}_{2} \nu_{\tau})$ & $0.012$ & $0.023$ \\
$\mathcal{B}(\snu \to \tilde{\chi}^{\pm} \tau^{\mp})$ & $0.026$ & 
$0.052$    \\
$\mathcal{B}(\snu \to \tilde{\tau}^{\mp} W^{\pm})$ & $0.620$ & $0.772$    
\end{tabular}
 \end{ruledtabular}
 \end{table}
The first point to notice is that the neutrino energy at resonance is 
approximately 300 and $500\gev$ for the two superparticle spectra under 
study---squarely in the atmospheric neutrino range. Accordingly, there is 
no need to invoke unknown mechanisms  to produce the required 
neutrino ``beam,'' as we must for the $\nu \bar{\nu} \to Z^{0}$ 
resonance. The modest resonant energies follow from the fact that the 
neutralino mass is more than eleven orders of magnitude larger than 
the (relic) neutrino mass. In both cases, inelastic 
processes---sneutrino decays that do not return to the entrance 
channel---are prominent.

There ends the good news, at least for the Universe at large.  Whereas
we expect that stable neutrinos should be the most abundant particles
in the Universe after the photons of the cosmic microwave background,
the neutralino gas is on average very tenuous.  The neutralino fraction
of the Universe (identified with the dark-matter fraction) is
\begin{equation}
    \Omega_{\neuter}h^{2} = \frac{\varrho_{\neuter}h^{2}}{\varrho_{c}} ,
\end{equation}
where the numerical value of the critical density  is
\begin{equation}
    \varrho_{c} = 1.05 \times 10^{-5}h^{2}\gev\cm^{-3}.
\end{equation}
Consequently the mass density of relic neutralinos is
\begin{equation}
    \varrho_{\neuter} = 1.05 \times 10^{-5}\cdot  \Omega_{\neuter}h^{2}\gev\cm^{-3} = 
    \bar{\mathcal{N}}_{\neuter}M(\neut),
\end{equation}
where $\bar{\mathcal{N}}_{\neuter}$ is the mean number density of relic 
neutralinos throughout the Universe.
In the two models we consider here, $\bar{\mathcal{N}}_{\neuter} \lesssim
10^{-8}\cm^{-3}$, some ten orders of magnitude smaller than the relic
neutrino density {and 31 orders of magnitude smaller than the 
density of electrons in water}.  

The peak cross sections  for inelastic sneutrino formation
\begin{equation}
    \sigma_{\snu:\mathrm{inel}}^{\mathrm{peak}} =
 \frac{8\pi m_{\snu}^{2}}{(m_{\snu}^{2}-M_{\neut}^{2})^{2}}\mathcal{B}(\snu
 \to \neut\nu)[1- \mathcal{B}(\snu
 \to \neut\nu)]\,, \label{eq:inelsig}
 \end{equation}
are about an order of
magnitude smaller than that~\cite{Barenboim:2004di} for $Z^{0}$ formation,
$\sigma(\nu\bar{\nu}\to Z)^{\mathrm{peak}}_{\mathrm{visible}} = 
0.80\,\sigma(\nu\bar{\nu}\to Z)^{\mathrm{peak}} =
365\nb$. The cross sections integrated over the peak are some 
 fifty to seventy times smaller than
$\Gamma_{Z}\sigma(\nu\bar{\nu}\to Z)^{\mathrm{peak}}_{\mathrm{visible}} = 912\nb\gev$,
where $\Gamma_{Z} = 2.4952\gev$.  
At the resonance peak, the interaction length for neutrinos on 
neutralinos in the present Universe is
\begin{equation}
    \mathcal{L}_{\mathrm{int}}^{\tilde{\nu}} = 
    1/\sigma_{\tilde{\nu}}^{\mathrm{peak}}\bar{\mathcal{N}}_{\neuter}
\end{equation}
For the two spectra under consideration, we find $ 
\mathcal{L}_{\mathrm{int}}^{\tilde{\nu}} \approx {(1.2,2.4)} \times 
10^{15}\mpc$ for model (I',L'), for $\nu_{e,\mu}$ interactions, with 
similar values for $\nu_{\tau}$ interactions. These values are roughly 11 orders of 
magnitude longer than the interaction length at resonance for $\nu \bar{\nu} \to 
Z^{0}$, which is $\mathcal{L}^{\nu\bar{\nu}}_{\mathrm{int}} \approx 1.2 \times 
10^{4}\mpc$ in the current Universe~\cite{Barenboim:2004di}. A 
distance of $10^{15}\mpc$ corresponds to approximately $10^{21.5}$ 
light years, so exceeds the $1.3 \times 10^{10}$-year age of the 
Universe by a prodigious factor. If $\nu \chi^{0}_{1} \to \snu$ 
coannihilation were to be observable, absorption lines would not be 
the signature! We should have to rely  on the direct detection of 
few-GeV $\gamma$ rays or charged particles from the inelastic decay 
chains that lead back to the $\nu\neut$ ground state.

\begin{figure}[tb]
   \centerline{\includegraphics[width=9cm]{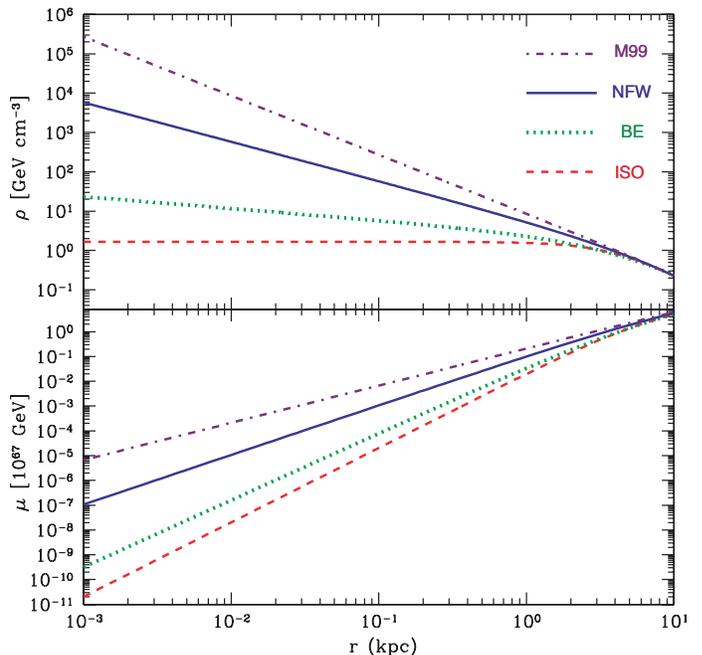}}
   \vspace*{-12pt}
\caption{Top panel: Dark-matter density profiles given by the 
form (\ref{eq:rho}) with parameters specified in 
Table~\ref{tab:Prof}. The solid (blue) curve shows the canonical 
Navarro--Frenk--White universal 
profile~\cite{Navarro:1996he}. The dot-dashed (purple) curve 
represents the cuspier profile of Moore and collaborators~\cite{Moore:1999nt}.
Also shown are the Binney--Evans profile~\cite{Binney:2001wu} (dotted
green line) and the nonsingular profile corresponding to an isothermal 
sphere (dashed red line). Bottom panel: The corresponding enclosed-mass 
profiles. 
}
\label{fig:galdens}
\end{figure}

Our location in the Universe may not be privileged, but it is not
average.  Cold dark matter clusters in galaxies.  
Most of the analytic expressions proposed in the literature for 
dark-matter halo profiles can be cast
in the form
\begin{equation}
    \varrho_{\neuter}(r)= \frac{ \varrho_0 }{
{(r/a)}^{\gamma}\,{\left[1+{(r/a)}^\alpha\right]}^{(\beta-\gamma)/\alpha}
}\;, \label{eq:rho} 
\end{equation}
where $\varrho_0$ is a characteristic
density, $a$ sets the radius of the halo, $\gamma$ is the asymptotic
logarithmic slope at the center, $\beta$ is the slope as
$r\to\infty$, and $\alpha$ controls the detailed shape of the profile in
the intermediate regions around $a$.
\begin{table}
\caption{Parameters of 
four radial density profiles of the Milky Way dark halo considered in
Ref.~\cite{Ascasibar:2005rw}, according to the parametrization given by
expression (\ref{eq:rho}), plus the mass enclosed in a sphere of 
radius $r_{\odot}$.  
\label{tab:Prof}}
\begin{center}
    \renewcommand{\arraystretch}{1.5}
\begin{ruledtabular}
    \begin{tabular}{ccccccc}
 Model & $\alpha$ & $\beta$ & $\gamma$ & $a$ [kpc] & 
 $\varrho_0$ [GeV cm$^{-3}$] & $\mu_{\odot}~[10^{67}\gev]$\\
\hline
 ISO & 2   & 2 &  0   &  4.0 & 1.655 & $3.8847$ \\
 BE~\cite{Binney:2001wu}  & 1   & 3 & 0.3  & 10.2 & 1.459 & $3.8351$ \\
 NFW \cite{Navarro:1996he} & 1   & 3 &  1   & 16.7 & 0.347 & $4.4205$ \\
 M99 \cite{Moore:1999nt} & 1.5 & 3 & 1.5  & 29.5 & 0.0536 & $4.8675$ \\
\end{tabular}
\end{ruledtabular}
\end{center}
\end{table}
A useful catalogue of dark-matter halo profiles for the Milky Way 
galaxy is given in Ref.~\cite{Ascasibar:2005rw}. The parameters of 
four profiles examined there are collected in Table~\ref{tab:Prof}. In each case, the density 
parameter $\varrho_{0}$ is adjusted to reproduce the presumed 
dark-matter density (neglecting local influences)
at our distance ($r_{\odot} = 8.5\kpc$) from the galactic center, 
$\varrho_{\neuter}(r_{\odot}) = 0.3\gev\cm^{-3}$~\cite{Bertone:2005xv}, 
which represents five orders of magnitude enhancement 
over the relic density in the Universe at large. The resulting 
neutralino number densities, $\mathcal{N}_{\tilde{\chi}}(r_{\odot}) 
\equiv \varrho_{\neuter}(r_{\odot}) / M(\neut)$, are given in 
Table~\ref{table:sugra}; the density 
profiles themselves are shown in Figure~\ref{fig:galdens}, along with 
the dark mass enclosed within a sphere of specified galactocentric radius.    

What little we know about the distribution of dark matter within the 
solar system has been deduced by investigation possible dark-matter 
perturbations on the orbits of planets and asteroids. In a recent 
contribution, Khriplovich and Pitjeva~\cite{Khriplovich:2006sq} have 
deduced bounds on the amount of dark matter that lie within the 
orbits of Mercury, Earth, and Mars. On the assumption that the 
dark-matter density is constant within the spherical volumes delimited 
by these three planetary orbits, the Earth and Mars orbits constrain
\begin{equation}
    \varrho_{\neuter}^{\odot} < 10^{-19}\g\cm^{-3} =
    5.6 \times 10^{4}\gev\cm^{-3} ,
    \label{eqn:ssdense}
\end{equation}
which would imply upper bounds on the number density of neutralinos at 
Earth, $\mathcal{N}_{\neuter}^{\oplus} \lesssim 350\cm^{-3}$.

What are the prospects for observing the 
coannihilation of neutrinos on the relic neutralino background? 
Consider first the interaction of neutrinos created in the collisions 
of cosmic rays with Earth's atmosphere. According to the calculations 
of the atmospheric neutrino flux reviewed by Gaisser \& 
Honda~\cite{Gaisser:2002jj}, 
the muon-neutrino flux at $E_{\nu} \approx 300\gev$ is
\begin{equation}
    \frac{dN_{\nu}}{dE_{\nu}} \approx 
    \frac{{10^{-5}}}{300}\cm^{-2}\s^{-1}\gev^{-1} .
\end{equation}
The quantity 
$\Gamma_{\tilde{\nu}}\,\sigma^{\mathrm{peak}}_{\tilde{\nu}:\mathrm{inel}}$ 
provides a rough measure of the coannihilation cross section 
integrated over the resonance peak. Consequently, the product
\begin{equation}
    \mathcal{R}_{\tilde{\nu}} \equiv
    \Gamma_{\tilde{\nu}}\,\sigma^{\mathrm{peak}}_{\tilde{\nu}:\mathrm{inel}} \cdot 
    \frac{dN_{\nu}}{dE_{\nu}}   
\end{equation}
characterizes the \snu\ coannihilation rate of a single neutralino in
the rain of atmospheric neutrinos. 

For model I', we compute 
the coannihilation rate $\mathcal{R}_{\snu}^{(\mathrm{I')}} \approx 
4.3 \times {10^{-40}}\s^{-1}$. If we consider the neutralino density 
for the Universe at large, then the rate per unit target volume is 
\begin{eqnarray}
    \hat{\mathcal{R}}_{\snu}^{(\mathrm{I')}} = 
    \bar{\mathcal{N}}_{\neuter} 
    \mathcal{R}_{\snu}^{(\mathrm{I')}}   &
    \approx & 3.3 \times 10^{-48}\cm^{-3}\s^{-1} \nonumber \\
    & \approx &  10^{-40}\cm^{-3}\y^{-1} , 
    \label{eqn:intrates}
\end{eqnarray}
which is laughably small, and  smaller still
for model L'. The rates (\ref{eqn:intrates}) are multiplied by 
$(2.9 \times 10^{5}, 5.2 \times 10^{10})$ if we consider the mean 
density at Earth's distance from the galactic center or the 
orbital-dynamics upper 
bound (\ref{eqn:ssdense}) on the density of dark matter in the solar system.

To pursue this line to a logical conclusion, we ask how many
interactions of atmospheric neutrinos with relic neutralinos occur in
the atmosphere per unit time.  Taking the Earth to be a sphere with
radius $R_{\oplus} = 6\,371\km$, we compute the volume of a 10-km-thick
shell above the Earth's surface to be $V_{\mathrm{atm}} = 5.1 \times
10^{24}\cm^{3}$.  The number of sneutrino events induced in this volume
is therefore $\lesssim 3 \times 10^{-5}\y^{-1}$, taking 
the largest conceivable neutralino number density.
Such an infinitesimal rate renders moot a discussion of
signatures and detection efficiencies. Setting aside questions of 
detection, the total number of events within the Earth's 
volume, $V_{\oplus} = 1.08 \times 10^{27}\cm^{3}$, is no more than $6
\times 10^{-3}\y^{-1}$.

Extraterrestrial sources may interact with relic neutrinos over a larger 
volume. A representative estimate~\cite{Stecker:2005hn} of the diffuse neutrino flux from 
active galactic nuclei suggests 
that, at $E_{\nu} \approx 300\gev$, 
the flux of cosmic neutrinos arriving from all directions is 
\begin{equation}
    \frac{dN_{\nu}}{dE_{\nu}} \approx 
    6.3 \times 10^{-17}\cm^{-2}\s^{-1}\gev^{-1} ,    
    \label{eq:floyd}
\end{equation}
approximately $2 \times 10^{{-9}} \times$ the vertical atmospheric 
neutrino flux we have just considered. The upper 
bound on the amount of dark matter in the sphere defined by Earth's 
orbit, $\mu(1\hbox{ au}) < 7.85 \times 
10^{44}\gev$~\cite{Khriplovich:2006sq}, could therefore be the site 
of up to (144,130) $\nu \neut \to \snu$ inelastic interactions per 
year for model (I',L'). 

Let us compute the contribution of each event,
at its own location, to the signal recorded by a detector of unit area
at Earth.  It will be sufficient to assume that the detector records
signals from all relevant directions with perfect efficiency.  We
define the vector from the Sun to the detector as $\vec{s}
\equiv (0, 0, s)$, and denote the location of the target as $\vec{r}
\equiv r(\sin\theta\cos\phi, \sin\theta\sin\phi, \cos\theta)$.  Then
the vector that points from the target to the detector is $\vec{d} =
\vec{s}- \vec{r}$, so that $d^{2} = r^{2} + s^{2} - 2rs\cos\theta$.
The decay products of the sneutrino produced at $\vec{r}$ are
distributed isotropically.  Accordingly, the fraction incident on a
detector of area $\mathcal{A}$ will be $f = \mathcal{A}/{4\pi d^{2}}$,
the fraction of the solid angle that the detector subtends.

The effective number of targets seen by a detector of unit area at 
$\vec{s}$ is given by
\begin{eqnarray}
    n_{\mathrm{eff}}(s) & \equiv  & \frac{1}{4\pi M_{\neuter}}
    \int d^{3}\mathbf{r}\;\frac{\varrho_{\neuter}(r)}{d^{2}} \label{eq:weighted} \\
    & = & \frac{{1}}{2M_{\neuter}} \int_{-1}^{1} \!\!d(\cos\theta)                                                                 
	      \int_{0}^{s}\frac{r^{2}\,dr\varrho_{\neuter}(r)}{r^{2} + s^{2} - 
	      2rs\cos\theta}  .  \nonumber
\end{eqnarray}
For the special case of a constant density 
$\varrho_{\neuter}^{\odot} = 5.6 \times 10^{4}\gev\cm^{-3}$, and $s = 
1\hbox{ au} = 1.496 \times 10^{13}\cm$, the 
effective number of targets is 
\begin{equation}
    n_{\mathrm{eff}}(s) = 
    \frac{\varrho_{\neuter}^{\odot}s}{2M_{\neuter}} \approx 3 \times 
    10^{15}\cm^{-2}.
    \label{eq:neffsolar}
\end{equation}
Using the cosmic-neutrino flux (\ref{eq:floyd}), we estimate that a 
detector in the vicinity of Earth would be sensitive to $7.7 \times 
10^{-26}\hbox{ events}\cm^{-2}\y^{-1}$.
With such a small number of events, there is no hope of detecting the 
few-GeV gamma rays that would be created in the cascade back to  
\neut.

The halo of our galaxy contains a great quantity of dark matter: for 
the NFW profile, $4.4 \times 10^{67}\gev$ lies within the galactocentric 
radius of our solar system. The number of neutralino targets with which 
neutrinos might coannihilate to form sneutrinos is thus $(3.1, 2.4) 
\times 10^{65}$ for model (I', L'). Focusing again on model I' and 
taking the estimate (\ref{eq:floyd}) for the cosmic-neutrino flux, we 
find that the sneutrino formation rate throughout the galaxy is $2.6 
\times 10^{17}\s^{-1} = 8.1 \times 10^{24}\y^{-1}$. These rates are 
prodigious in absolute terms, but represent an insignificant disturbance to the 
galaxy's neutralino population.

Can we hope to observe the sneutrino formation that might be bubbling
away in our neighborhood?  in this case, we
define $\vec{s}\equiv (0, 0, s)$ to be the vector from the galactic center 
to the detector, and use (\ref{eq:weighted}) to compute the effective number 
of targets seen from Earth.
For the NFW profile, we find $n_{\mathrm{eff}}(r_{\odot}) = 4.9 \times 
10^{19}\cm^{-2}$. For a detector of any plausible area, only
a tiny fraction of the $3.1 \times 10^{65}$ 
neutralinos in the halo produce a visible signal. Indeed, using the 
neutrino flux (\ref{eq:floyd}), we expect that a detector in the 
vicinity of Earth would be sensitive to $1.3 \times 10^{-21}\hbox{ 
events}\cm^{-2}\y^{-1}$.

\begin{figure}[tb]
\centerline{\includegraphics[width=9cm]{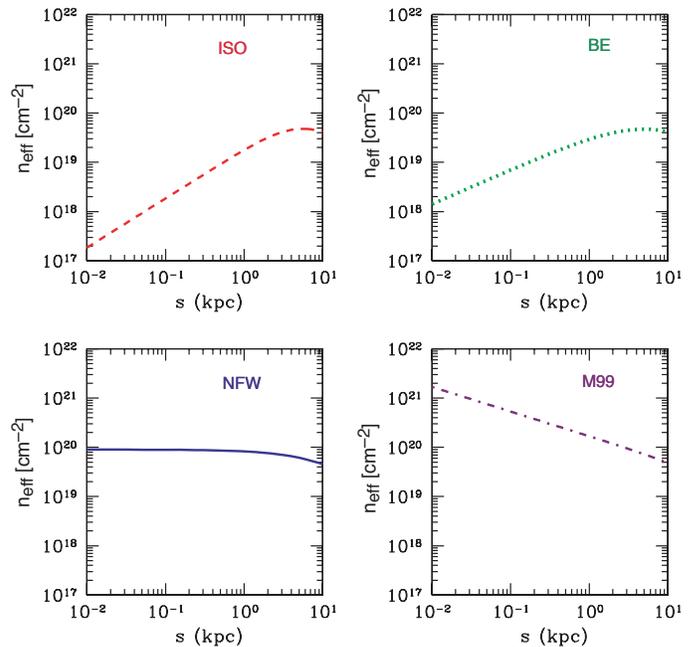}}
\vspace*{-6pt}
\caption{ Effective number of neutralino targets per unit of detector 
area viewed from a 
location at radius $s$ from the galactic center, according to four 
dark-matter density profiles for the Milky Way galaxy. Our location 
corresponds to $s = 8.5\kpc$.} \label{fig:where}
\end{figure}

It is not obvious that Earth's location should 
happen to be optimal for viewing coannihilation events in the galaxy. In 
Figure~\ref{fig:where} we show how the effective number of targets 
viewed depends on $s$, the galactocentric radius of the 
observation point.
For the smooth ISO and BE profiles, our solar system lies near the optimal 
distance, while for the NFW profile the effective number of targets 
viewed is insensitive to the detector position. The most singular 
profile we consider, M99, favors an observatory close to the galactic 
center---not that we could contemplate one---but even in this case, 
the sensitivity is enhanced by less than two orders of magnitude, 
which is far too little to enable detection. Singular profiles have 
more effect on the rates for neutralino-neutralino annihilations, 
which are proportional to the square of the neutralino density.

If neutralinos account for much of the cold dark matter of the 
Universe, then it is possible that neutrino--neutralino 
coannihilation into sneutrinos is happening all around us, but at a 
rate that will forever elude detection.

\begin{acknowledgments}
We thank Luis Iba\~{n}ez for posing a question that stimulated this
investigation.  We are grateful to C\'{e}line Boehm, Dan Hooper, and
Peter Skands for helpful conversations.  Fermilab is operated by
Universities Research Association Inc.\ under Contract No.\
DE-AC02-76CH03000 with the U.S.\ Department of Energy.  GB and CQ
acknowledge the hospitality of the CERN Particle Theory Group.

\end{acknowledgments}


\begin{thebibliography}{99}


\bibitem{Weiler:1982qy}
  T.~J.~Weiler,
  Phys.\ Rev.\ Lett.\  {\bf 49}, 234 (1982).
  
  \bibitem{Eberle:2004ua}
    B.~Eberle, A.~Ringwald, L.~Song and T.~J.~Weiler,
    Phys.\ Rev.\ D {\bf 70}, 023007 (2004)
    [arXiv:hep-ph/0401203].
    
    \bibitem{Barenboim:2004di}
      G.~Barenboim, O.~Mena Requejo and C.~Quigg,
      Phys.\ Rev.\ D {\bf 71}, 083002 (2005)
      [arXiv:hep-ph/0412122].

\bibitem{Spergel:2006hy}
D.~N.~Spergel {\it et al.},
``Wilkinson Microwave Anisotropy Probe (WMAP) three year results:
Implications for cosmology,''
arXiv:astro-ph/0603449.
      
\bibitem{PDGcosmo}
O.~Lahav and A.~R.~Liddle, ``The Cosmological Parameters,'' in
\textit{Review of Particle Properties,} Ref.~\cite{Eidelman:2004wy},
\S21.

\bibitem{Eidelman:2004wy}
S.~Eidelman {\it et al.}  [Particle Data Group],
Phys.\ Lett.\ B {\bf 592}, 1 (2004), and 2005 partial update for 
edition 2006, \url{http://pdg.lbl.gov}.

\bibitem{Bertone:2004pz}
  G.~Bertone, D.~Hooper and J.~Silk,
  Phys.\ Rept.\  {\bf 405}, 279 (2005)
  [arXiv:hep-ph/0404175].
  
  \bibitem{Bernabei:2003za}
    R.~Bernabei {\it et al.},
    Riv.\ Nuovo Cim.\  {\bf 26N1}, 1 (2003)
    [arXiv:astro-ph/0307403].

\bibitem{Akerib:2005kh}
  D.~S.~Akerib {\it et al.}  [CDMS Collaboration],
  Phys.\ Rev.\ Lett.\ {\bf 96,} 011302 (2006) [arXiv:astro-ph/0509259];
Phys.\ Rev.\ D {\bf 73,} 011102(R) (2006)
    [arXiv:astro-ph/0509269].
   
    \bibitem{Sanglard:2005we}
    V.~Sanglard {\it et al.}  [The EDELWEISS Collaboration],
    Phys.\ Rev.\ D {\bf 71}, 122002 (2005)
    [arXiv:astro-ph/0503265].

\bibitem{Battaglia:2003ab}
  M.~Battaglia, A.~De Roeck, J.~R.~Ellis, F.~Gianotti, K.~A.~Olive and L.~Pape,
  Eur.\ Phys.\ J.\ C {\bf 33}, 273 (2004)
  [arXiv:hep-ph/0306219].
 
    \bibitem{Datta:2004sr}
      A.~Datta, D.~Fargion and B.~Mele,
      JHEP {\bf 0509}, 007 (2005)
      [arXiv:hep-ph/0410176].
      
      \bibitem{Porod:2003um}
	W.~Porod,
	``\textsf{SPheno}, a program for calculating supersymmetric spectra, SUSY particle
	decays and SUSY particle production at $e^{+} e^{-}$ colliders,''
	Comput.\ Phys.\ Commun.\  {\bf 153}, 275 (2003)
	[arXiv:hep-ph/0301101].
      
	\bibitem{Belanger:2004yn}
	  G.~B\'{e}langer, F.~Boudjema, A.~Pukhov and A.~Semenov,
	  ``\textsf{MicrOMEGAs}: Version 1.3,''
	  Comput.\ Phys.\ Commun.\  {\bf 174}, 577 (2006)
	  [arXiv:hep-ph/0405253].
      
      \bibitem{Belanger:2005jk}
	G.~B\'{e}langer, S.~Kraml and A.~Pukhov,
	Phys.\ Rev.\ D {\bf 72}, 015003 (2005)
	[arXiv:hep-ph/0502079];
	  B.~C.~Allanach, S.~Kraml and W.~Porod,
	  JHEP {\bf 0303}, 016 (2003)
	  [arXiv:hep-ph/0302102];
    \textit{Comparison of SUSY spectrum generators:
     mass spectra, relic densities, etc.,} \url{http://cern.ch/kraml/comparison/}.
     
     \bibitem{Group:2006qt}
       Tevatron Electroweak Working Group,
       ``Combination of CDF and D\O\ results on the mass of the top quark,''
       arXiv:hep-ex/0603039.
       
\bibitem{Ascasibar:2005rw}
 Y.~Ascasibar, P.~Jean, C.~Boehm and J.~Knoedlseder,
 ``Constraints on dark matter and the shape of the Milky Way dark halo from
 the 511-keV line,'' 
 arXiv:astro-ph/0507142.

 \bibitem{Binney:2001wu}
   J.~J.~Binney and N.~W.~Evans,
   Mon.\ Not.\ Roy.\ Astron.\ Soc.\  {\bf 327}, L27 (2001)
   [arXiv:astro-ph/0108505].
   
\bibitem{Navarro:1996he}
J.~F.~Navarro, C.~S.~Frenk and S.~D.~M.~White,
Astrophys.\ J.\  {\bf 490}, 493 (1997).

\bibitem{Moore:1999nt}
  B.~Moore, et al., 
  Astrophys.\ J.\  {\bf 524}, L19 (1999).



\bibitem{Bertone:2005xv}
  G.~Bertone and D.~Merritt,
  Mod.\ Phys.\ Lett.\ A {\bf 20}, 1021 (2005)
  [arXiv:astro-ph/0504422].

  \bibitem{Khriplovich:2006sq}
    I.~B.~Khriplovich and E.~V.~Pitjeva,
    ``Upper limits on density of dark matter in Solar system,''
    arXiv:astro-ph/0601422.
See also
      L.~Iorio,
      ``Solar System planetary orbital motions and dark matter,''
      arXiv:gr-qc/0602095.

    \bibitem{Gaisser:2002jj}
      T.~K.~Gaisser and M.~Honda,
      Ann.\ Rev.\ Nucl.\ Part.\ Sci.\  {\bf 52}, 153 (2002)
      [arXiv:hep-ph/0203272].
      

\bibitem{Stecker:2005hn}
  F.~W.~Stecker,
  Phys.\ Rev.\ D {\bf 72}, 107301 (2005)
  [arXiv:astro-ph/0510537].
\end{thebibliography}
\end{document}